\documentclass[twoside]{article}
\usepackage{pmart,fleqn,graphicx,amsmath}

\begin{document}

\title{The Overlapping Muffin-Tin Approximation}
\author{M. Zwierzycki$^{a,b}$ and O.K. Andersen$^{b}$}
\address{$^{a}$Institute of Molecular Physics, Polish Academy of
  Sciences, \\  Smoluchowskiego 17, 60-179 Pozna\'n, Poland\\ 
  $^{b}$Max Planck Institute for Solid State Research, 
  Heisenbergstrasse 1,\\ 70569 Stuttgart, Germany
}
\date{\today}
\maketitle

\begin{abstract}
We present the formalism and demonstrate the use of the overlapping
muffin-tin approximation (OMTA). This fits a full potential to a
superposition of spherically symmetric short-ranged potential wells plus a
constant. For one-electron potentials of this form, the standard
multiple-scattering methods can solve Schr\"{o}dingers' equation correctly
to 1st order in the potential overlap. Choosing an augmented-plane-wave
method as the source of the full potential, we illustrate the procedure for
diamond-structured Si. First, we compare the potential in the Si-centered
OMTA with the full potential, and then compare the corresponding OMTA $N$-th
order muffin-tin orbital and full-potential LAPW band structures. We find
that the two latter agree qualitatively for a wide range of overlaps and
that the valence bands have an rms deviation of 20 meV/electron for 30\%
radial overlap. Smaller overlaps give worse potentials and larger overlaps
give larger 2nd-order errors of the multiple-scattering method. To further
remove the mean error of the bands for small overlaps is simple.
\end{abstract}

%
%

%
%

\section{Introduction}

The linear muffin-tin orbital (LMTO) method in the atomic-spheres
approximation (ASA) \cite{Andersen75,Skriver} has been used for over three
decades as an \emph{ab initio} method for electronic-structure calculations.
One of the defining characteristics of this method is its small, localized 
\cite{TBLMTO} basis set which provides highly efficient computation and does
not require crystalline symmetry \cite{Turek}. Equally important is the fact
that the similarity of LMTOs to atomic orbitals makes the method physically
transparent. This means that it is easy to extract \emph{e.g.} information
about the nature of chemical bonds, the symmetry of band states, a.s.o..

The extreme speed and simplicity of the LMTO-ASA is to a large extent due to
the ASA which takes the one-electron potential and charge density to be
spherically symmetric inside space-filling Wigner-Seitz spheres whose
overlap is neglected. This approximation is very good, not only for
closely-packed solids but also for open systems in which the interstital
sites have high symmetry and can be taken as centers for "empty"
Wigner-Seitz spheres containing no protons but only electrons. The diamond
structure is such a case, because both its potential and charge density are
are approximately spherical inside equal-sized spheres packed on a
body-centered cubic (bcc) lattice. However, for open systems with atoms of
different sizes and/or with low symmetry, it can be cumbersome, if not
impossible to find a good ASA.

Since most systems of current interest are low-dimensional and
multi-component, and since the cost of computation has followed Moors' law
for decades, the LMTO-ASA method is now mainly used to extract information
about the electrons from accurate calculations which use one of the
plane-wave supercell methods. These methods employ the variational principle
with large plane-wave basis sets, either for the full-potential Schr\"{o}%
dinger equation with linear augmented plane waves (LAPWs) \cite{Andersen75,
Singh:94} or projector-augmented plane waves (PAWs) \cite{PAW}, or for a
pseudopotential with straight plane waves \cite{Payne:92}. In this mode of
operation, one adjusts the LMTO-ASA sphere sizes and empty-sphere positions
by trial and error such as to obtain the best possible agreement with the
band structure obtained with an accurate method.

Attempts \emph{have} been made to improve the accuracy while preserving the
desirable features of the LMTO-ASA. First \cite{MTOpointOfView}, it was
realized that the multiple-scattering method of Korringa, Kohn and Rostoker
(KKR) \cite{KKR}, the mother of all MTO methods, solves Schr\"{o}dingers'
equation not only for a MT potential, that is a potential which is
spherically symmetric inside non-overlapping spheres and flat ($\equiv $0%
\textrm{)} in between, but also for a \emph{superposition} of spherically
symmetric potential-wells to 1st order in their overlap, with \emph{no}
change of the formalism. This explained why the ASA had worked much better
than the MT approximation (MTA) and begged for the use of even larger
potential spheres. The overlapping muffin-tin approximation (OMTA), however,
turned out \emph{not} to work with conventional LMTOs because the
energy-dependence of their envelope functions, the decaying Hankel
functions, had not been linearized away in the same way as the
energy-dependence of the augmenting partial waves inside the spheres, but
put to a constant (e.g.$\,$=$0$)$.$ As a consequence, \emph{exact} MTOs
(EMTOs) with envelopes of energy-dependent, screened solutions of the wave
equation had to be found, and then linearized \cite{Trieste}. With the
resulting, so-called 3rd-generation LMTOs (LMTO3s), it became possible to
solve Schr\"{o}dingers' equation for diamond-structured silicon using the
OMTA without empty spheres \cite{Andersen:98}. Specifically, the ASA
potential calculated self-consistently \emph{with} empty spheres (and the
14\% radial overlap of bcc WS spheres) was least-squares' fitted to a
superposition of potential wells centered only on the Si sites, whereafter
Schr\"{o}dingers' equation was solved for this OMTA potential with the LMTO3
method. By varying the OMTA overlap, a minimum rms error of 80 meV per
valence-band electron was found around 30\% overlap: for smaller overlaps
the OMTA-fit was worse, and for larger overlaps the $2$nd-order overlap
error of the LMTO3 method dominated. This ability of the LMTO3 method to
compute the band structure for an OMTA potential was not developed into a
charge-selfconsistent method, despite several attempts \cite%
{RashmiCatia,unpubl}, but for the EMTO Greens'-function method this was done
by Vitos et al. who devised a spherical-cell approximation for
closely-packed alloys and demonstrated its efficiency for overlaps around $%
30\%$ \cite{EMTO}. Instead, the unique ability of the LMTO3 method to
generate super-minimal basis sets through downfolding was explored, but that
led to the next problem: The smaller the basis set, the longer the range of
each MTO, and the stronger its energy dependence. For very small basis sets,
this energy dependence may be so strong that linearization is insufficient.
That problem was finally solved through the development of a polynomial
approximation of general order $(N)$ in the MTO Hilbert space, and the
resulting $N$MTO method \cite{Andersen:2000} has now been in use for nearly
ten years, in particular for generating few-orbital, low-energy Hamiltonians
in studies of correlated electron systems, but always in connection with
potentials generated by the LMTO-ASA.

Here, we shall present and apply the least-squares' formalism to fit a full
potential in the form delivered by any augmented plane-wave method to the
OMTA form. This procedure is demonstrated for diamond-structured silicon,
first by comparison of the full with the OMTA potential, and then by
comparison of the full-potential LAPW and OMTA $N$MTO band structures. Our
current strategy for extracting electronic information from an accurate
plane-wave calculation, is thus to fit its output potential to the OMTA and
then perform $N$MTO calculations for the OMTA potential. This procedure is
more simple and accurate than going via the LMTO-ASA.

\section{The Overlapping Muffin-Tin Approximation}

In this section we shall explain the least-squares' procedure for obtaining
the OMTA to a full potential. In the first two subsections we shall present
the general formulas \cite{unpubl}, and in the last we shall present results
for the specific analytic form of the full potential provided by an
augmented plane-wave method.

We wish to minimize the value of mean squared deviation,%
\begin{equation}
\Delta F=\frac{1}{\Omega }\int_{\Omega }\left[ F(\mathbf{r})-\left(
g+\sum\nolimits_{j}f_{j}(|\mathbf{r}-\mathbf{R}_{j}|)\right) \right] ^{2}d%
\mathbf{r,}  \label{eq:mean_sq_dev}
\end{equation}%
of the the full potential, $F\left( \mathbf{r}\right) ,$ from its OMTA
counterpart, $g+\sum\nolimits_{j}f_{j}(r_{j})$. The spherically symmetric potential
wells $%
f_{j}(r_{j})$ vanish for $r_{j}>s_{j}$ and are centered at the sites
$%
\mathbf{R}_{j}$ of the atoms. $\Omega $ is the (cell) volume. To be
determined are thus the radial functions $f_{j}\left( r\right) $ and
the constant $g,$ for given $F(\mathbf{r}),$ $\Omega ,$
$\mathbf{R}_{j},$ and $%
s_{j}$.

\subsection{Determining the spherical potential wells, $f_{j}(r)$}

\label{sec:f-equation} Variation of $\Delta F$ with respect to $f_{j}(r)$
leads to the equation: 
\begin{align}
0& =\frac{\delta }{\delta f_{j}(r)}\Delta F=\frac{1}{4\pi r^{2}}\int_{\Omega
}\delta (|\mathbf{r}-\mathbf{R}_{j}|-r)\left( F(\mathbf{r}%
)-g-\sum\nolimits_{j^{\prime }}f_{j^{\prime }}(|\mathbf{r}-\mathbf{R}%
_{j^{\prime }}|)\right) d\mathbf{r}  \notag \\
& \equiv \theta (s_{j}-r)\left( \bar{F}^{j,r}-g\right) -f_{j}(r)-\theta
(s_{j}-r)\sum\nolimits_{j^{\prime }\neq j}\bar{f}_{j^{\prime }}^{j,r}
\label{eq:func_diff_f}
\end{align}%
where $\bar{F}^{j,r}$ and $\bar{f}_{j^{\prime }}^{j,r}$ are the averages of
the respective functions over the sphere centered at site $j$ and of radius $%
r$. Eq.~(\ref{eq:func_diff_f}) thus translates into the requirement that the
spherical averages of $F(\mathbf{r})-g$ and $\sum_{j^{\prime }}f_{j^{\prime
}}(r_{j^{\prime }})$ around site $j$ be the same. The problem of calculating 
$\bar{F}^{j,r}$ for the LAPW potential will be addressed in Sec.~\ref%
{sec:flapw_sph_avg}. By reordering of the terms in Eq.~(\ref{eq:func_diff_f}%
), and evaluating the spherical averages, we obtain:%
\begin{equation}
rf_{j}(r)=\theta (s_{j}-r)\left[ r\left( \bar{F}^{j,r}-g\right)
-\sum_{j^{\prime }\neq j}\frac{\theta (s_{j}+s_{j^{\prime }}-d_{jj^{\prime
}})}{2d_{jj^{\prime }}}\int_{d_{jj^{\prime }}-r}^{s_{j^{\prime }}}r^{\prime
}f_{j^{\prime }}(r^{\prime })dr^{\prime }\right] ,  \label{eq:f_equation}
\end{equation}%
where $d_{jj^{\prime }}=|\mathbf{R}_{j}-\mathbf{R}_{j^{\prime }}|$ is the
distance between respective sites. We have assumed that $s_{j^{\prime
}}<d_{jj^{\prime }},$ i.e. that a potential sphere never includes the center
of another sphere (moderate overlap). Repeating the procedure for all
inequivalent sites, Eq.s (\ref{eq:f_equation}) become a set of\emph{\ linear
integral equations} which may be solved for a given $g=g_{0}$ by
self-consistent iteration. The iteration can be initiated\textit{\ e.g.} by
taking $f_{j^{\prime }}(r^{\prime })=\theta (s_{j^{\prime }}-r^{\prime
})\left( \bar{F}^{j^{\prime },r^{\prime }}-g\right) $ on the right-hand side
of Eq.$\,$(\ref{eq:f_equation}).

\subsection{Determining the constant background, $g$}

\label{sec:g-equation} A similar expression is obtained for the optimal
value of the constant $g$, namely: 
\begin{equation}
0=\frac{\partial }{\partial g}\Delta F=\bar{F}-g-\sum\nolimits_{j}\bar{f}_{j}
\label{eq:g_opt}
\end{equation}%
where $\bar{F}$ and $\bar{f}_{j}$ are the volume averages of the respective
functions: 
\begin{equation}
\bar{F}=\frac{1}{\Omega }\int_{\Omega }F(\mathbf{r})d\mathbf{r,}\quad 
\mathrm{and}\quad \bar{f}_{j}=\frac{1}{\Omega }\int_{\Omega }f_{j}(|\mathbf{r%
}-\mathbf{R}_{j}|)d\mathbf{r}=\frac{4\pi }{\Omega }%
\int_{0}^{s_{j}}f_{j}(r)r^{2}dr  \label{eq:Ff_vol_avg}
\end{equation}%
Just like before, Eq.~(\ref{eq:g_opt}) expresses the condition of equality
of averages of the full and OMTA potentials.

We need to solve the \emph{coupled} $f$ and $g$ equations. Since these are
linear, we may use the superposition principle to obtain:%
\begin{equation*}
f_{j}\left( g;r\right) =f_{j}\left( g_{0};r\right) -\left( g-g_{0}\right)
h_{j}\left( r\right) ,
\end{equation*}%
where the "structure functions" $h_{j}\left( r\right) $ are the solutions of
the $f$-equations (\ref{eq:f_equation}) for $F\left( \mathbf{r}\right)
-g\equiv 1.$ The optimal $g$-value is thus:%
\begin{equation*}
g=g_{0}+\left( \bar{F}-g_{0}+\sum\nolimits_{j}\bar{f}_{j}\left( g_{0}\right)
\right) \left/ \left( 1-\sum\nolimits_{j}\bar{h}_{j}\right) \right. .
\end{equation*}

\subsection{Averages of the LAPW full potential}

\label{sec:flapw_sph_avg} The procedure outlined in section \ref%
{sec:f-equation} requires calculation of the volume average $\bar{F},$ and
the spherical averages, $\bar{F}^{j,r},$ for $r\leq s_{j}<\min_{j^{\prime
}}d_{jj^{\prime }}$ around all inequivalent sites of the full potential.
This task can be achieved with relative ease thanks to the analytical form
of the potential delivered by an augmented plane-wave method: Space is
divided into a set of atom-centered, \emph{non}-overlapping augmentation
spheres with radii, $a_{j},$ and the interstitial. Inside an augmentation
sphere the full potential is given by the spherical-harmonics $\left(
Y_{lm}\right) $ expansion:%
\begin{equation}
F(\mathbf{r})=\sum_{lm}F_{jlm}(r_{j})Y_{lm}(\hat{\mathbf{r}}_{j})\;\;\mathrm{%
for}\;\;r_{j}<a_{j}  \label{eq:flapw_mt}
\end{equation}%
where $\mathbf{r}_{j}\equiv \mathbf{r}-\mathbf{R}_{j}\equiv r_{j}\,\hat{%
\mathbf{r}}_{j}.$ In the interstitial, the full potential is described as a
linear combination of plane waves which we shall let extend throughout space:%
\begin{eqnarray}
F(\mathbf{r}) &=&\sum_{j}\theta \left( a_{j}-r_{j}\right) \sum_{lm}\left(
F_{jlm}(r_{j})-\sum\nolimits_{G}F_{jlm}\left( G\right)
\,j_{l}(Gr_{j})\right) Y_{lm}(\hat{\mathbf{r}}_{j})  \notag \\
&&+\sum\nolimits_{\mathbf{G}}F\left( \mathbf{G}\right) e^{i\mathbf{G}\cdot 
\mathbf{r}}.  \label{FP}
\end{eqnarray}%
The first term here provides the augmentation of the plane waves, i.e. it
ads expression (\ref{eq:flapw_mt}) and subtracts the corresponding
plane-wave part inside the spheres. $\mathbf{G}\equiv G\mathbf{\hat{G}}$ are
the reciprocal lattice vectors and $F\left( -\mathbf{G}\right) =F^{\ast
}\left( \mathbf{G}\right) $ because $F\left( \mathbf{r}\right) $ is real.
Moreover, $j_{l}$ are the spherical Bessel functions, and 
\begin{equation}
F_{jlm}\left( G\right) \equiv 4\pi i^{l}\sum\nolimits_{\mathbf{\hat{G}}%
}F\left( \mathbf{G}\right) e^{i\mathbf{G}\cdot \mathbf{R}_{j}}Y_{lm}^{\ast }(%
\hat{\mathbf{G}}).  \label{FG}
\end{equation}

The spherical average around site $j$ is the spherical component of the
potential (\ref{eq:flapw_mt}) inside the augmentation sphere: $\bar{F}%
^{j,r}=F_{j00}\left( r\right) /\sqrt{4\pi },$ if $r<a_{j}.$ If $r>a_{j}$ and
the $r$-sphere at site $j$ cuts into the augmentation-sphere at site $%
j^{\prime },$ the spherical average of the corresponding term in the first
line of Eq.$\,$(\ref{FP}) is most easily evaluated when, for the
spherical-harmonics expansion inside the $j^{\prime }$-sphere, we turn the $%
z^{\prime }$-axis to point towards $\mathbf{R}_{j},$ i.e. along the $\mathbf{%
\hat{d}}_{jj^{\prime }}$ direction:%
\begin{equation*}
Y_{lm}(\hat{\mathbf{r}}_{j^{\prime }})=\sum\nolimits_{m^{\prime
}}D_{mm^{\prime }}^{l}\left( \mathbf{\hat{d}}_{jj^{\prime }}\right)
Y_{lm^{\prime }}\left( \hat{\mathbf{r}}_{j^{\prime }}\cdot \mathbf{\hat{d}}%
_{jj^{\prime }}\right) \equiv \sum\nolimits_{m^{\prime }}D_{mm^{\prime
}}^{l}\left( \mathbf{\hat{d}}_{jj^{\prime }}\right) Y_{lm^{\prime }}(\theta
^{\prime },\varphi ^{\prime }),
\end{equation*}%
because then the integral over $\varphi ^{\prime }$ selects the term with $%
Y_{l0}\left( \theta ^{\prime },\varphi ^{\prime }\right) =\sqrt{\frac{2l+1}{%
4\pi }}P_{l}\left( \cos \theta ^{\prime }\right) $ and only the integral
over $\theta ^{\prime }$ remains. Including now also the spherical average
of the second line in Eq.$\,$(\ref{FP}), the result becomes:%
\begin{eqnarray}
\bar{F}^{j,r} &=&\sum_{j^{\prime }\neq j}\frac{\theta \left( d_{jj^{\prime
}}-r\right) }{2rd_{jj^{\prime }}}\sum_{l}\int_{d_{jj^{\prime
}}-r}^{a_{j^{\prime }}}r^{\prime }F_{jj^{\prime }l}\left( r^{\prime }\right)
P_{l}\left( \frac{r^{\prime 2}-r^{2}+d_{jj^{\prime }}^{2}}{2r^{\prime
}d_{jj^{\prime }}}\right) dr^{\prime }  \notag \\
&&+\sum_{G}j_{0}\left( Gr\right) F_{j00}\left( G\right) /\sqrt{4\pi }\,,
\label{Favg}
\end{eqnarray}%
where $j_{0}\left( x\right) =\sin x/x$ and%
\begin{gather*}
F_{jj^{\prime }l}\left( r^{\prime }\right) \equiv \theta \left( a_{j^{\prime
}}-r^{\prime }\right) \sqrt{\frac{2l+1}{4\pi }}\sum_{m}\left[ F_{j^{\prime
}lm}(r^{\prime })-\sum\nolimits_{G}F_{j^{\prime }lm}\left( G\right)
\,j_{l}(Gr^{\prime })\right] D_{m0}^{l}\left( \mathbf{\hat{d}}_{jj^{\prime
}}\right)  \\
=\theta \left( a_{j^{\prime }}-r^{\prime }\right) \sqrt{\frac{2l+1}{4\pi }}%
\left[ \sum_{m}F_{j^{\prime }lm}(r^{\prime })D_{m0}^{l}\left( \mathbf{\hat{d}%
}_{jj^{\prime }}\right) -\sum_{G}F_{j^{\prime }l0}^{j}\left( G\right)
\,j_{l}(Gr^{\prime })\right] .
\end{gather*}%
Here, $F_{j^{\prime }l0}^{j}\left( G\right) $ is given by Eq.$\,$(\ref{FG})
with $Y_{l0}^{\ast }(\hat{\mathbf{G}})$ substituted by $\sqrt{\frac{2l+1}{%
4\pi }}P_{l}\left( \hat{\mathbf{G}}\mathbf{\cdot \hat{d}}_{jj^{\prime
}}\right) .$

Finally, the volume average of the full potential (\ref{FP}) is%
\begin{equation*}
\bar{F}=\frac{\sqrt{4\pi }}{\Omega }\sum_{j}\left(
\int_{0}^{a_{j}}F_{j00}(r)r^{2}dr-a_{j}^{3}\sum_{G}\frac{j_{1}\left(
Ga_{j}\right) }{Ga_{j}}F_{j00}\left( G\right) \right) +F\left( \mathbf{G}%
\mathrm{=}\mathbf{0}\right) ,
\end{equation*}%
where $j_{1}\left( x\right) =\left( \sin x-x\cos x\right) /x^{2}.$

\begin{figure}[t]
\includegraphics[width=\textwidth,clip=true]{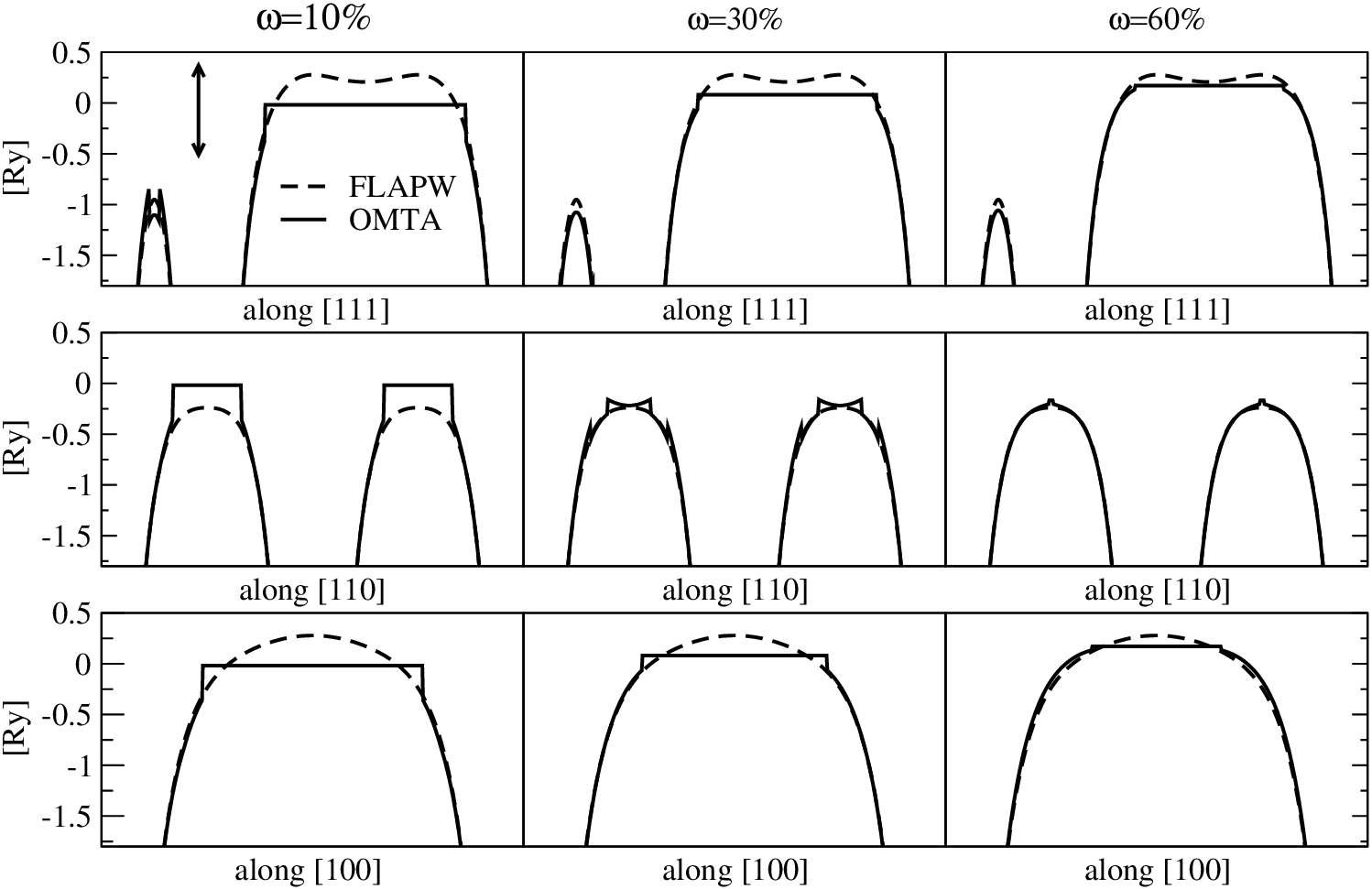}
\caption{Full potential for diamond-structured Si along the main
crystallographic axes (dashed) and its OMTAs for 10, 30, and 60\% radial
overlaps (full). The double-headed arrow indicates the range of the valence
bands.}
\label{fig:pot_fig}
\end{figure}

\section{Example: Diamond-structured silicon}

In this section we shall compare the potential and band structure calculated
self-consistently using the full potential LAPW method as implemented in
WIEN2k \cite{wien2k} with respectively the OMTA to this full potential and
its band structure calculated using the $N$MTO-OMTA method. As in the
previous (AE)SA-\textit{vs}-OMTA LMTO3 study \cite{Andersen:98}, in the
present full-\textit{vs}-OMTA $N$MTO study we choose diamond-structured Si
and OMTAs with atom-centered spherical wells in order that the study be
relevant for open structures in general.

Fig.~\ref{fig:pot_fig} compares the full potential with its OMTA for radial
overlaps, $\omega \equiv (2s-d)/d,$ of 10, 30, and 60\%. We see a marked
improvement in the quality of the fit as the overlap, i.e. the range of $%
f\left( r\right) ,$ increases. With $60\%$ overlap the OMTA follows the full
potential not only inside the potential wells, \textit{i.e.} close to the
nuclei where the spherical approximation is at its best, but also along the
bond between the Si atoms and in the interstitial space.

\begin{figure}[t]
\centering  \includegraphics[width=0.5\textwidth,clip=true]{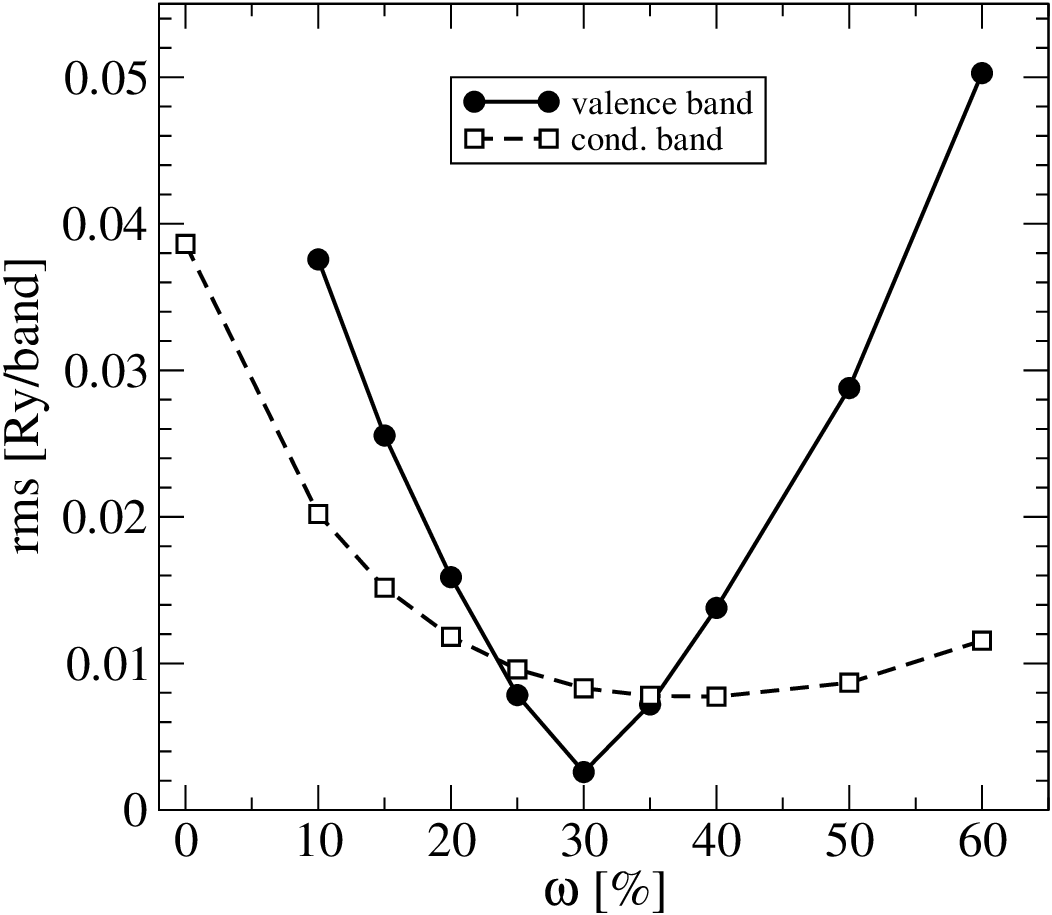}
\caption{Rms errors of the $N$MTO-OMTA valence and four lowest conduction
bands as functions of the radial overlap, with respect to the full-potential
LAPW bands.}
\label{fig:rms}
\end{figure}

From this, one might believe that it is always preferential to use large
overlaps. But this is not so, because multiple-scattering methods are only
"self-correcting" to 1st-order in the potential overlap ($\propto \omega
^{2})$ \cite{MTOpointOfView,Andersen:98}. For large overlaps, the errors of
higher order ($\propto \omega ^{4}$ and higher) outweigh the effect of the
improved description of the potential. We thus expect that there exists an
optimal value of the overlap which yields the best accuracy of the
calculated band states. In Fig.~2 this is confirmed by explicit calculations
of root mean squared (rms) error of the $N$MTO-OMTA bands with respect to
their full-potential LAPW counterparts as a function of the radial overlap.
The two curves are for respectively the four valence and the four lowest
conduction bands. For both types of bands, the error initially decreases as
a result of the improved OMTA. For the valence bands the optimum of 3$~$%
mRy/band = 20~meV/electron is reached at $30\%$ overlap, after which the
error increases sharply, whereas for the conduction bands the optimum of
50~meV/e is reached at $40\%$, wherafter the multiple-scattering errors
increase slowly. This difference in behaviour is due to the valence states
being bonding and the conduction states antibonding, whereby the former
probe the overlap region more than the latter. Whereas the rms-error curve
for the conduction band is nicely flat, that of the valence band is \emph{%
not.}

\begin{figure}[t]
\centering  \includegraphics[width=\textwidth,clip=true]{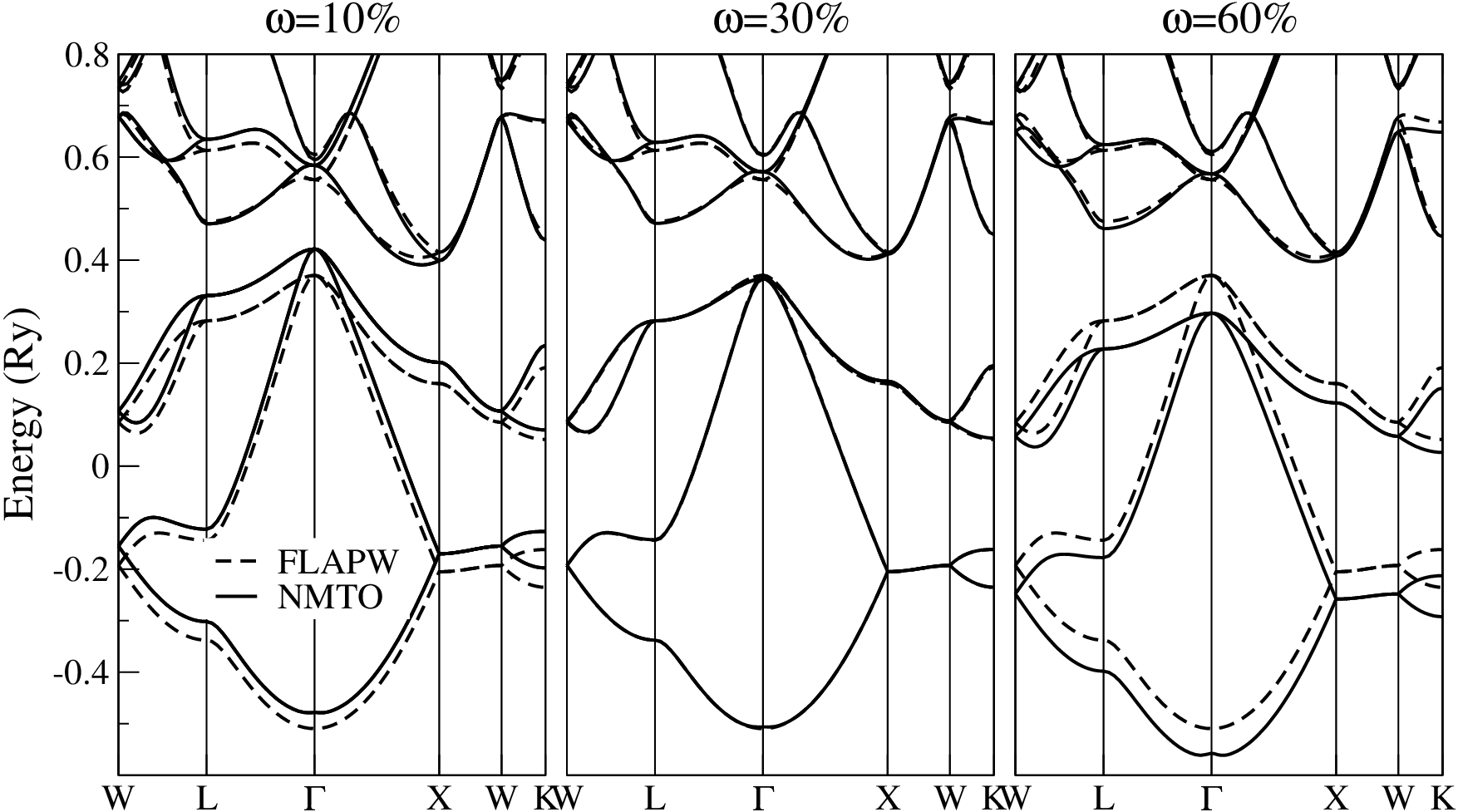}
\caption{Band structure of Si calculated with the full potential LAPW method
(dashed) and with the $N$MTO-OMTA method for $10$, 30, and 60\% radial
overlap (full). The $N$MTO basis set consisted of the 9 $s$, $p,$ and $d$ Si-%
$N$MTOs with the $f$s downfolded and had $N$=2 with expansion energies -0.5,
0.3, and 0.6~Ry \protect\cite{Andersen:2000}.}
\label{fig:bands}
\end{figure}

Fig.~\ref{fig:bands} shows the full-potential LAPW band structure and the $N$%
MTO band structures for the three OMTA potentials in Fig.~\ref{fig:pot_fig}.
Excellent agreement is achieved with $30\%$ overlap for both conduction and
valence bands. For smaller overlaps, the $N$MTO-OMTA bands lie to high, the
valence bands in particular, and for larger overlaps they lie too low. The
first point follows from the variational principle and the second from the
fact that the multiple-scattering error $\propto \omega ^{4}$ is negative
definite. This was explained in the previous study \cite{Andersen:98}, which
also demonstrated the use of various corrections. The simplest of these is
to remove the \emph{mean} error of the energies for a group of bands, say
the valence bands, by weighting the squared deviation in Eq.$\,$\ref%
{eq:mean_sq_dev} with the charge density of those bands, and then keep the
weighting in the $g$-equation only (see Sect.~\ref{sec:g-equation}). Since
the LAPW valence charge density has the same form as the full potential,
implementation of this requires use of formulas given in Sect.~\ref%
{sec:flapw_sph_avg}. This correction will move the $N$MTO-OMTA bands down by
the appropriate amount for small overlaps and, hence, considerably lower the
rms-error curve for small overlaps. An elegant correction of the
large-overlap errors remains to be found; the straight-forward but clumsy
one is to evaluate the kinetic energy properly in the overlap region for the 
$N$MTO basis.

\end{document}